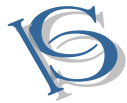

IJCSBI.ORG

# The Computer-Linguistic Analysis of Socio-Demographic Profile of Virtual Community Member

**Yuriy Syerov, Andriy Peleschyshyn, Solomia Fedushko**
Social Communications and Information Activities Department,
Lviv Polytechnic National University,
Ukraine, Lviv, S. Bandery street 12,

## ABSTRACT

This article considers the current problem of investigation and development of computer-linguistic analysis of socio-demographic profile of virtual community member. Web-members' socio-demographic characteristics' profile validation based on analysis of socio-demographic characteristics. The topicality of the paper is determined by the necessity to identify the web-community member by means of computer-linguistic analysis of their information track. The formal model of basic socio-demographic characteristics of virtual communities' member is formed. The structural model of lingvo-communicative indicators of socio-demographic characteristics of the web-members and common algorithm of the formation of lingvo-communicative indicators based on processing training sample are developed. Types of the computer-linguistic analysis of indicative characteristics are studied and classifications of lingvo-communicative indicators of gender, age and sphere of activities of web-community member is established. Also, the formal model of the basic socio-demographic characteristics of web-communities' member is introduced.

## 1. INTRODUCTION

The computer-linguistic analysis of socio-demographic analysis of the web-community member consists in the analysis of member's information track. Nowadays, developing the method of personal data validation [1, 2] of the maximum amount of information about virtual community member is an important issue in the fields of social communication and internet-marketing. To achieve the aim, the following research tasks should be fulfilled: to form the model of basic socio-demographic characteristics of virtual communities' member; to develop the structural model of indicators of socio-demographic characteristics of the web-community members and common algorithm of the formation of lingvo-communicative indicators based on processing training sample; to study the types of the computer-linguistic analysis of indicative characteristics; to establish classifications of lingvo-communicative indicators; to introduce the formal model of the basic characteristics of virtual communities' member.

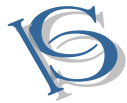

## 2. BACKGROUND STUDY

Sets of lingvo-communicative indicators for verification of the socio-demographic characteristics of web-community's members were forming by experts. The base of this research was studding scientific theories and the ideology of the leading scientists.

A new introduction to the study of the relation between gender and language use are investigations of the leading experts in the fields: gender differences in the level of speech key phrases [3], modern approaches to the study of language and gender [4], sociolinguistic analysis of discussions of conferences [5], comparison relations style of gender speech and world perception [6], gender research of language in online communication [3] [7] [8], comparison of managers language [9], experimental investigations of conversations of married couples [10], gender language research in three communities with different social relations [11].

A modern approach to the study of language and age are investigations of the outstanding scientists in the fields: web-communication of adolescents [12], sociolect of adolescents [13], methods and means of internet-communications of adolescents [7], teen style [14].

Studies of language and sphere of activities are investigations of the leading scientists in the fields: Physico-Mathematical, Technical and Economic sphere [16], Chemicals sphere [17] [18], Sociological, Historical, Philosophical and Political sphere [19] [20], Natural sphere [21], Medical sphere [22], Philological-Pedagogical sphere [23], Sphere of Architecture and Art [24], Sphere of Physical Training and Sport [25], Agricultural sphere [26],Military sphere [27], Legal sphere [28] [29]

## 3. STRUCTURAL MODEL OF LINGVO-COMMUNICATIVE INDICATORS OF SOCIO-DEMOGRAPHIC CHARACTERISTICS

The structural model of lingvo-communicative indicators of socio-demographic characteristics of the web-community members is consisting of four levels. Figure 1 shows that socio-demographic characteristics are determined by set values. Each of studied characteristics has two determinate values. By-turn, one of the values of socio-demographic characteristics necessarily is assigned to each web-community members.

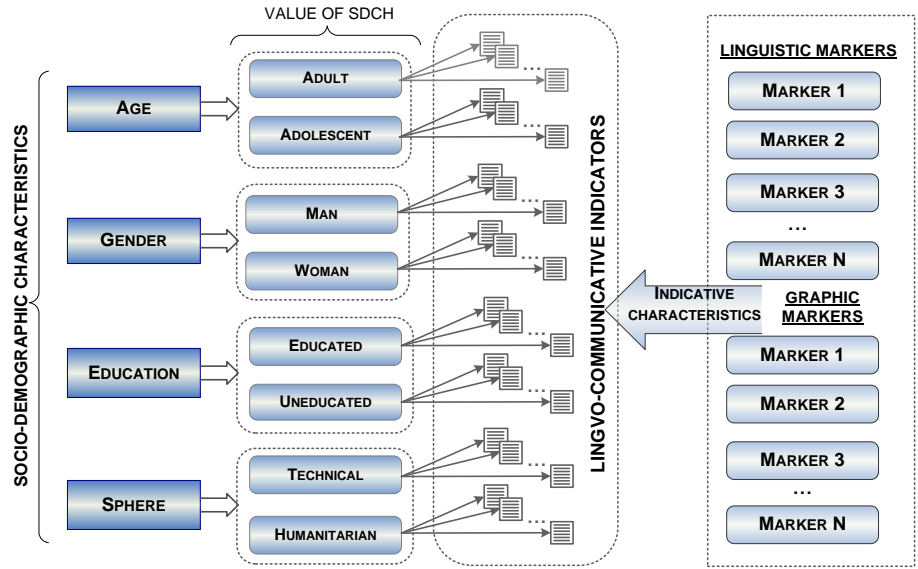

**Figure 1. Scheme of the structural model of lingvo-communicative indicators of socio-demographic characteristics of the web-community members**

Every one of level components of the structural model of lingvo-communicative indicators of socio-demographic characteristics of the web-community members will consider further in this study.

## 3.1  Notion of socio-demographic characteristics

The socio-demographic characteristics [30] are necessary to consider because these characteristics are base for formation of socio-demographic profiles of web-community's member. In social communication the socio-demographic characteristics are defined as a set of basic characteristics of the account in web community. Analysis of socio-demographic characteristics of virtual community's member lies in the forming socio-demographic profile's [31] of web-community's member. The socio-demographic characteristics are provided by member of this web-community.

In general, the socio-demographic characteristics include: age, gender, material security, position, marital status, education level, profession and work experience, employment, location, religious and political views, etc.

The notion of socio-demographic characteristics has a wide range of use in many sciences. Socio-demographic characteristics are defined in various fields as important parameters [32].

Every socio-demographic characteristic of virtual community's member is determining by analysis of linguistic features (markers) in virtual community's members communication. Thus, the socio-demographic characteristics are a set of social estimation criteria and important parameters of human activity.

### 3.1.1 Definition of lingvo-communicative indicators

Lingvo-communicative indicators are special features of language and communication of online community's members, which can be traced in his information track. The lingvo-communicative indicators determine the community member belonging to a particular set of socio-demographic characteristics. Lingvo-communicative indicator of socio-demographic characteristics of virtual community's members is a set of linguistics and graphic features that are inherent to a web-communication specific online community member. These indicators establish identity virtual member to the set of socio-demographic characteristics and determine the value of socio-demographic characteristics, actually. That is, the person in the course of communicative activity uses features that helps expert to explore gender and age identity, level of education, field of interest. For example, smiles, lexical and graphic signs, etc.

### 3.1.2 Definition of markers

In view of the fact that linguistic and communication indicators are set linguistic and graphic markers that define certain socio-demographic characteristics of online community members using computer-linguistic processing of the virtual communities content. The definition of the marker looks like this explanation: marker is linguistic or graphic feature, which contains information about the socio-demographic beginning of anonymous web-member and identifies the authenticated web-community member and group of web-community member and their socio-demographic identity. Thus, markers are features of online-communication of web-member that in information track are traced. According to the type of content markers are divided into: graphical and linguistic markers. Linguistic marker is a language feature (word, phrase or sentence) of internet-communication that indicates belonging of the author of web-community content to certain socio-demographic characteristic. Graphic marker is a graphic figure (avatars or visual identification, user bars, emotions, etc.) that indicates belonging of the author of web-community content to certain socio-demographic characteristic. Vector of markers by method of computer-linguistic processing of virtual web-community member information track is obtained.

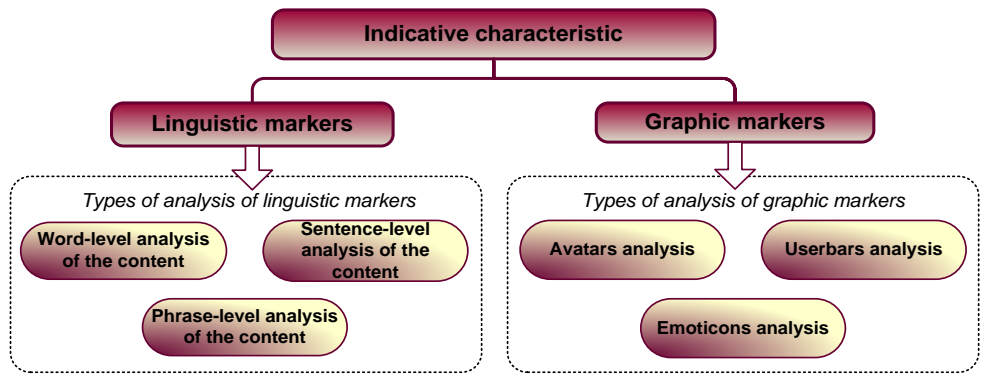

**Figure 2. Types of the computer-linguistic analysis of indicative characteristics**

Types of the computer-linguistic analysis of indicative characteristics in the Figure 2 are showed.

### 3.2 Common algorithm of the formation of lingvo-communicative indicators based on processing training sample

The common algorithm of the formation of lingvo-communicative indicators based on processing training sample consists of the following stages (See Figure 3):

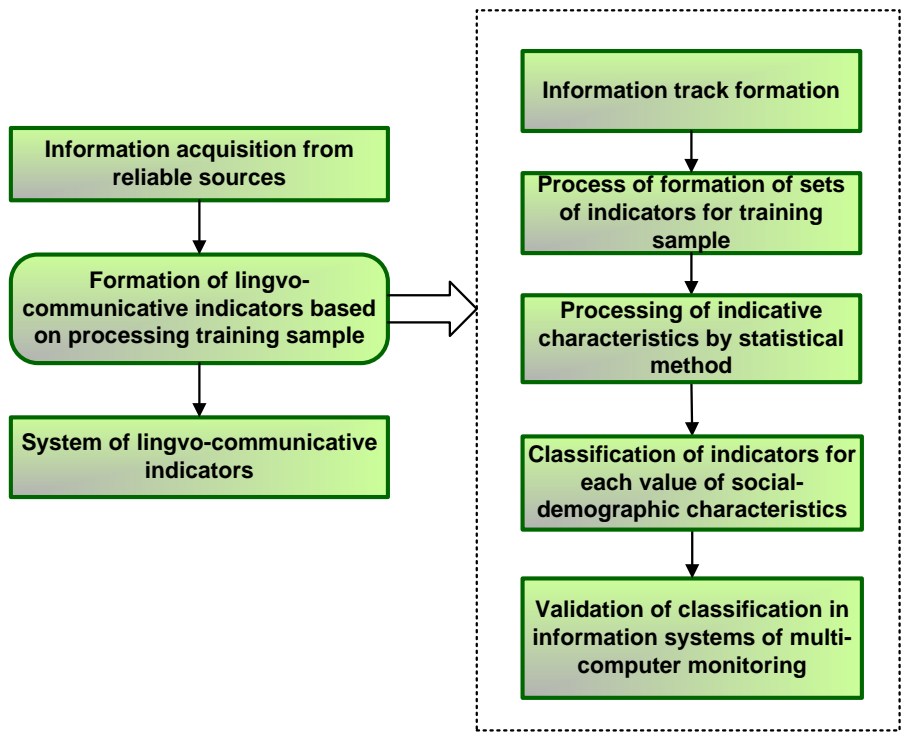

**Figure 3. Scheme of common algorithm of the formation of lingvo-communicative indicators based on processing training sample**

Currently the most important stages of the algorithm will be reviewed:

### 3.2.1 System of lingvo-communicative indicators

System of lingvo-communicative indicators is the basis for the software development for the personal information validation of virtual community members by means of "Socio-demographic characteristics verifier".

### 3.2.2 Processing of indicative characteristics by statistical method

In this stage of algorithm implemented automated statistical data analysis using application packages that ensure factor, cluster and discriminant analysis.

### 3.2.3 Validation of classification in information systems of multi-computer monitoring

Verification of algorithm results carried out by using automated data-processing monitoring system.

### 3.2.4 Process of formation of sets of indicators for training sample

Process of formation of sets of indicators for training sample consists in three stages.

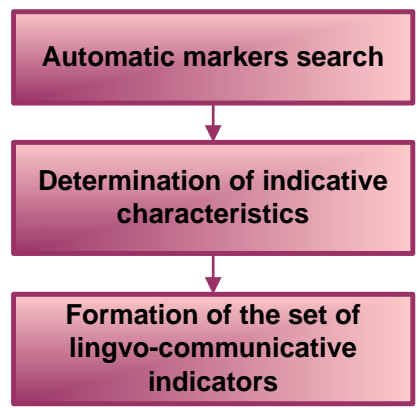

**Figure 4. Scheme of formation of lingvo-communicative indicators based on processing training sample**

Figure 4 shows the steps of the process of formation of sets of indicators for training sample.

## 3.3 Formation of lingvo-communicative indicators based on processing training sample

The moderators and administrators must accurately track socio-demographic characteristic identity of web-community members in order to prevent conflicts in the virtual community. The method of determining socio-demographic characteristic of web-community members to improve the management of virtual communities is used. Context of messages and themes of discussion significantly effect on the result of this research. This fact in the study is taken into consideration and diversified training sample of users' messages of all web-forums topics is realized. The discussions in the threads of these forums are equally considered. All these discussions in connection with a variety of interests and passions of men and women, adults and adolescents are aroused.

*3.3.1 Formation of gender lingvo-communicative indicators*

The analysis of a set of gender linguistic characteristics of virtual community members describe as follows:

$$\text{Gend}(U_i) = \left(\text{Gend}_j(U_i)\right)_{j=1}^{N_i^{\text{Gend}}} \ (1)$$

where $\left(\text{Gend}_j(U_i)\right)_{j=1}^{N_i^{\text{Gend}}}$ – set of gender lingvo-communicative indicator of web-community member; $N_i^{\text{Gend}}$ – number of certain gender lingvo-communicative indicator of web-community member.

The classification of lingvo-communicative indicators of gender is shown in Table 1.

Table 1. Classification of lingvo-communicative indicators of gender

| Indicator of gender | Denotation |
|---|---|
| The emotional component | Gender-A |
| Cultural aspects | Gender-B |
| References | Gender-C |
| Guidelines and instructions | Gender-D |
| Lexical aspect | Gender-E |
| Method of expressing content | Gender-F |
| Timeframe | Gender-G |
| Insignificance | Gender-H |
| Power, influence and authoritativeness | Gender-I |
| Beneficiation of language | Gender-J |
| Composition | Gender-K |
| Concretization | Gender-L |

*3.3.2 Formation of age lingvo-communicative indicators*

The analysis of a set of age linguistic characteristics of virtual community members describe as follows:

$$\text{Age}(U_i) = \left(\text{Age}_j(U_i)\right)_{j=1}^{N_i^{\text{Age}}} \ (2)$$

where $\left(\text{Age}_j(U_i)\right)_{j=1}^{N_i^{\text{Age}}}$ – set of age lingvo-communicative indicator of web-community member; $N_i^{\text{Age}}$ – number of certain age lingvo-communicative indicator of web-community member, $U_i$.

Table 2 shows classification of lingvo-communicative indicators of age.

**Table 2. Classification of lingvo-communicative indicators of age**

| Indicator of age | Denotation |
|---|---|
| Affiliative and aggressive style | Age-A |
| Slang variation | Age-B |
| Modulation of voice and sound similarity | Age-C |
| Text economy | Age-D |
| Non-codified units and non-verbal means | Age-E |
| Deformalization | Age-F |

*3.3.3 Formation of sphere of activity of lingvo-communicative indicators*

The analysis of a set of sphere linguistic characteristics of virtual community members describe as follows:

$$\text{Sphere}(U_i) = \left(\text{Sphere}_j(U_i)\right)_{j=1}^{N_i^{\text{Sphere}}} \quad (3)$$

where $\left(\text{Sphere}_j(U_i)\right)_{j=1}^{N_i^{\text{Sphere}}}$ – set of sphere lingvo-communicative indicator of web-community member; $N_i^{\text{Sphere}}$ – number of certain sphere lingvo-communicative indicator of web-community member.

The classification of lingvo-communicative indicators of sphere of activity is offered in Table 3.

**Table 3. Classification of lingvo-communicative indicators of sphere of activity**

| Indicator of sphere of activity | Denotation |
|---|---|
| Physico-Mathematical, Technical and Economic sphere Monday | Sphere-A |
| Chemicals sphere | Sphere-B |
| Sociological, Historical, Philosophical and Political sphere | Sphere-C |
| Natural sphere | Sphere-D |
| Medical sphere | Sphere-E |
| Philological-Pedagogical sphere | Sphere-F |
| Sphere of Architecture and Art | Sphere-G |
| Sphere of Physical Training and Sport | Sphere-H |
| Agricultural sphere | Sphere-I |
| Legal sphere | Sphere-J |
| Military sphere | Sphere-K |

### 3.4 A common model of lingvo-communicative indicators of socio-demographic characteristics of the web-community members

In this part of paper the formal model of the basic socio-demographic characteristics of virtual communities' member is introduced. Our socio-demographic characteristics model expresses the basic socio-demographic characteristics as ordered set of reliable socio-demographic characteristics of virtual community's member, and is described by a mathematical equation. The socio-demographic characteristics model includes only basic socio-demographic characteristics of virtual communities' member, such as "age", "sphere of activity" and "gender".

As previously was mentioned, researchers defined the information track as a set of all personal data of virtual community's member and the results of his communicative activity - the content, which is created by web member. The socio-demographic characteristics model of member virtual community – socio-demographic profile – describe as follows:

$$\text{SDP}\left(U^*\right) = \left(\text{SDCh}_j\left(U^*\right)\right)_{j=1}^{N^{\text{SDCh}(U^*)}} \quad (4)$$

where $\text{SDCh}_j\left(U^*\right)_{j=1}^{N^{\text{SDCh}(U^*)}}$ – ordered set of socio-demographic characteristics of community member $U^*$; $N^{\text{SDCh}}(U^*)$ – quantity of these characteristics of member $U^*$.

In the particular case, a set of socio-demographic characteristics is described as:

$$\text{SDCh}\left(U^*\right) = \left(\text{age}\left(U^*\right), \text{edu}\left(U^*\right), \text{gend}\left(U^*\right), \text{sphere}\left(U^*\right)\right), \quad (5)$$

where $\text{SDCh}(U^*)$ – a set of socio-demographic characteristics, $U^*$ – web-community member, $\text{age}(U^*)$ – a set of age-indicator of web-community member, $\text{edu}(U^*)$ – a set of education-indicator of web-community member, $\text{gend}(U^*)$ – a set of gender-indicator of web-community member, $\text{sphere}(U^*)$ – a set of sphere-indicator of web-community member.

For the convenience of the analysis of a set of linguistic and communicative indicators of age, gender and sphere of activity a virtual community member is describe as follows:

$$\text{Ind}\left(IO_i\right) = \left(\text{Ind}_j\left(IO_i\right)\right)_{j=1}^{N_i^{\text{Ind}(IO)}} \quad (6)$$

where $\left(\text{Ind}_j\left(IO_i\right)\right)_{j=1}^{N_i^{\text{Ind}(IO)}}$ – vector of lingvo-communicative indicator of web-community member, IO – indicative characteristics, Ind– lingvo-communicative indicator, $N_i^{Ind^{(IO)}}$ – number of these lingvo-communicative indicator of web-community member.

The value of *"i"* belongs to the set of natural numbers ($i \in N$) for investigated socio-demographic characteristics. However,

1) socio-demographic characteristic "gender" $1 \le i \le 12$, because this socio-demographic characteristicis determined twelve lingvo-communicative indicators.

2) socio-demographic characteristic "age" is determined six lingvo-communicative indicators, so $1 \le i \le 6$.

3) socio-demographic characteristic "sphere of activity" is determined eleven lingvo-communicative indicators, so $1 \le i \le 11$.

The monoatomic indicative characteristics of lingvo-communicative indicator of web-community member determine the specific markers, weight of markers and regulations of use:

$$IO_i = \langle M_i, W, R \rangle \qquad (7)$$

where $\langle M_i, W, R \rangle$ – vector of specific features, $M_i$ – marker, $W$ – weight of markers, $R$ – regulations of use.On the other part, the atomic indicator is defined as:

$$Ind_k = \left\{ IO_j \right\}_{j \in NI_k} (8)$$

where $\left\{ IO_j \right\}_{j \in NI_k}$ – the set of indicative characteristics, $IO_j$ – j-th indicative characteristics, $NI_k$ – set of numbers of indicative characteristics that indicative set is included (or formed).

Each of the indicative characteristic is described by vector of markers:

$$IO = \left\{ Marker_i \right\}_{i=1}^{N^{(Mr)}} (9)$$

where $\left\{ Marker_i \right\}_{i=1}^{N^{(Mr)}}$ – the set of markers, $Marker_i$ – i-th marker, $N^{(Mr)}$ – number of elements in the set $Marker$.

The socio-demographic characteristic of web-forum member is a set of pairs of markers and weights of markers. Weighting factor (weight of markers) is a factor of expression measures of certain marker of socio-demographic characteristics in the track information of web-community member. Weighting factor define the importance of marker for certain lingvo-communicative indicator of socio-demographic characteristic:

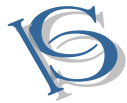

$$IO_i = \left\langle \left( Marker_j, v_{ij} \right) \right\rangle_{j=1}^{N^{(M_{IO_i})}} \quad (10)$$

where $Marker_j \in Marker$ – marker from a set of markers; $v_{ij}$ – weighting factor $Marker_j$ of certain indicative characteristic $IO_i$; $N^{(MIO_i)}$ – number of markers of indicative characteristic $IO_i$.

Markers of web-community content (is created by user $User_i$) are automatically detected using specialize software. This data form a set of markers of certain indicator of socio-demographic characteristic of web-member $MarkerList(User_i)$.

$$MarkerList(User_i) \subseteq Marker \quad (11)$$

Based on information track of web-community member the extent of congruence $User_i$ with indicative characteristic $IO_p$ defined by:

$$\mu\left(User_i, IO_p\right) = \frac{\sum_{j=1}^{N^{(MUser_i)}} v_{pj} n_{ij}}{\sum_{j=1}^{N^{(KMIIO_p)}} v_{pj}} \quad (12)$$

where $v_{pj} = v\left(IO_p, Marker_j\right)$ – weighting factor $Marker_j$ of indicative characteristic $IO_p$, $n_{ij} = n\left(Content_i, Marker_j\right)$ – number of markers using $Marker_j$ in content of web-community member $Content_i$.

During the analysis only just significant socio-demographic characteristic for proposed model of socio-demographic profile of web-community member are investigated (including age, gender and sphere of activity).

## 4. CONCLUSIONS

A question of urgent importance in the web-forum management and moderation is the development of a new approach to data verification which gives community members when they are resisted. The formal model of basic socio-demographic characteristics of virtual communities' member is formed. The structural model of lingvo-communicative indicators of socio-demographic characteristics of the web-community members and common algorithm of the formation of lingvo-communicative indicators based on processing training sample are developed. Types of the computer-linguistic analysis of indicative characteristics are studied and classifications of lingvo-communicative indicators of gender, age and sphere of activities of web-community member are established. Also, the formal model of the basic socio-demographic characteristics of virtual communities' member is introduced. As the result of analysis the verifiable personal information web-community members and its socio-demographic profile are obtained. So, the issue of socio-demographic profile of virtual community's member verifying is investigated. The socio-demographic profile of virtual community's member helps automatically administrated and monitored the web-forums. Thus, this paper presents a new approach to developing computer-linguistic method of socio-demographic profile formation.